# Observational tracking of the 2D structure of coronal mass ejections between the Sun and 1 AU

Running Heads: Tracking the 2D CME structure

N P Savani[1,*], J A Davies[2], C J Davis[2,4], D Shiota[3], , A P Rouillard[4,11], M J Owens[5,10] K Kusano[6, 12], V Bothmer[7], S P Bamford[8], C J Lintott[9], A Smith[9]

1 University Corporation for Atmospheric Physics (UCAR), Boulder, Co, USA
2 RAL Space, STFC Rutherford Appleton Laboratory, Harwell Oxford, Didcot, OX11 0QX, UK
3 Computational Astrophysics Laboratory, Advanced Science Institute, RIKEN, 2-1, Hirosawa, Wako, Saitama 351-0198, Japan
4 College of Science, George Mason University, Fairfax, VA 22030, USA
5 Space Environment Physic Group, University of Reading, Earley Gate, P.O. Box 243, Reading RG6 6BB, UK
6 Solar-Terrestrial Environment Laboratory, Nagoya University, Nagoya 464–8601, Japan
7 Göttingen University, Institut für Astrophysik, Göttingen, Germany
8 Centre for Astronomy and Particle Theory, University of Nottingham, University Park, Nottingham NG7 2RD, UK
9 Astrophysics, University of Oxford, Denys Wilkinson Building, Keble Road, Oxford OX1 3RH, UK
10 The Blackett Laboratory, Imperial College London, Prince Consort Road, London SW7 2BZ, UK
11 Space Science Division, Naval Research Laboratory, Washington, DC 20375-5352, USA
12 Japan Agency for Marine-Earth Science and Technology, Yokohama, Kanagawa 236-0001, Japan

Email: neel.savani02@imperial.ac.uk

**Abstract**

The Solar Terrestrial Relations Observatory (STEREO) provides high cadence and high resolution images of the structure and morphology of coronal mass ejections (CMEs) in the inner heliosphere. CME directions and propagation speeds have often been estimated through the use of time-elongation maps obtained from the STEREO Heliospheric Imager (HI) data. Many of these CMEs have been identified by citizen scientists working within the SolarStormWatch project (www.solarstormwatch.com) as they work towards providing robust real-time identification of Earth-directed CMEs. The wide field of view of HI allows scientists to directly observe the 2-dimensional (2D) structures, while the relative simplicity of time-elongation analysis means that it can be easily applied to many such events, thereby enabling a much deeper understanding of how CMEs evolve between the Sun and the Earth. For events with certain orientations, both the rear and front edge of the CME can be monitored at varying heliocentric distances (R) between the Sun and 1 AU. Here we take four example events with measurable position angle widths and identified by the citizen scientists. These events were chosen for the clarity of their structure within the HI cameras and their long track lengths in the time-elongation maps. We show a linear dependency with R for the growth of the radial width (W) and the 2D aspect ratio ($\chi$) of these CMEs, which are measured out to ~0.7AU. We derived a radial width estimate from a linear best fit for the average of the four CMEs; the equation followed W=0.14R + 0.04. The aspect ratio followed $\chi$=2.5R + 0.86.

# 1. Introduction

For many decades now, coronal mass ejections (CMEs) have been detected in situ and observed remotely (e.g. Hudson *et al.* 2006; Zurbuchen & Richardson 2006). Properties such as shape, size and speed have received particular interest (e.g. St Cyr *et al.* 2000; Chen 2011), but more so for those events that are directed towards Earth. The geomagnetic effects of interplanetary CMEs on Earth may vary from being severe in some cases to negligible in others (e.g. Kikuchi *et al.* 1978; Kikuchi *et al.* 2008). Predicting the geomagnetic impact of a CME and estimating how long their effects will persist is of interest to a diverse range of business sectors, e.g. from satellite telecommunications to national electrical infrastructure (Baker *et al.* 2008).

For many years, interplanetary scintillation (IPS) has provided scientists with indirect measurements of the changing structure of the solar wind and CMEs between the Sun and terrestrial distances (e.g. Watanabe & Schwenn 1989; Fujiki et al. 2005; Tokumaru et al. 2006; Tokumaru et al. 2010; Tokumaru et al. 2011). These studies have formed the foundations for certain 3D reconstruction techniques (Jackson *et al.* 2007; Bisi *et al.* 2010). During the last decade, spacecraft carrying wide-angle heliospheric imaging instruments have enabled scientists to routinely monitor the global 2D structure of CMEs from above the solar surface to terrestrial distances. The field of view of the solar mass-ejection imager (SMEI) on the Coriolis spacecraft (Eyles *et al.* 2003; Jackson *et al.* 2004) is limited to elongations beyond 18°, while the Heliospheric Imager (HI) instrument on STEREO (Eyles *et al.* 2009) is not orbiting Earth and therefore also able to observe CMEs travelling along the Sun-Earth line and thus enables scientists to directly compare the remote signatures to the in situ measurements (e.g. Davis et al. 2009; Lynch et al. 2010). The elongation angle is defined as the angle between the Sun and the object as seen by the observer.

The STEREO mission, launched in 2006 (Kaiser *et al.* 2008), consists of two spacecraft that follow a trajectory similar to that of the Earth. As they separate from each other at a rate of ~45° per year, one spacecraft travels ahead of Earth (ST-A) while the other lags behind (ST-B). Each spacecraft carries the Sun-Earth Connection Coronal and Heliospheric Investigation (SECCHI, Howard *et al.* 2008) imaging package; which contains an Extreme Ultraviolet

Imager (EUVI), two coronagraphs (COR-1 and COR-2), and the Heliospheric Imager (HI). The HI instrument on each STEREO spacecraft is made up of two wide-field visible-light imagers, HI-1 and HI-2 (Eyles *et al.* 2009). The fields of view of HI-1 and HI-2 are of 20° and 70° angular extent, respectively, and under ordinary operation are nominally centred at 13.7° and 53.4° elongation in the ecliptic plane. Thus the ecliptic plane corresponds to a horizontal line that runs through the centre of the fields of view.

To analyse data from these wide-field cameras, new methods have been derived which need no longer rely on assumptions of plane of sky propagation, traditionally applied to coronagraph data. At large distances from the Sun, the elongation angle ($\alpha$), is more appropriate than estimates of the physical distance derived from the plane of the sky approximation. Originally developed for LASCO coronagraph, Sheeley et al. (1999) estimated how the elongation angle of a transient propagating outwards from the Sun would vary with time. The authors deduced that a single plasma element propagating in a fixed direction to the Sun-spacecraft line ($\beta$) at a constant radial speed ($V_r$) would result in a unique time-elongation profile. Therefore $\beta$ is defined to the east of the Sun-spacecraft line for ST-A, and to the west for ST-B. Therefore by analysis of the time-elongation profile of a solar transient, an analytical best fit for $\beta$ and $V_r$ can be retrieved. This technique is usually referred to as the fixed-phi (FP) approximation, and has been extensively used for the analysis of time-elongation profiles from the STEREO HI instrument (e.g. Rouillard *et al.* 2008; Davies *et al.* 2009; Davis *et al.* 2009; Rouillard *et al.* 2009b; Savani *et al.* 2010). Another method called the point-P approximation was developed (Kahler & Webb 2007) in an attempt to take into account the Thomson sphere geometry (Vourlidas & Howard 2006). However results from the point-P approximation are poor for structures that are azimuthally narrow at 1AU, such as dense features entrained in co-interacting regions (CIRs, Rouillard *et al.* 2008; Rouillard *et al.* 2009b). Lugaz et al. (2009) developed a different method to improve on the limitations of both the fixed-phi and point-P technique. The authors assume that, in the plane of interest (usually the ecliptic), a CME can be modelled as a circle, with its rear edge anchored to the center of the Sun, and whose center propagates along a fixed radial trajectory. This method is often labelled the harmonic mean (HM) approximation, because the distance of the CME corresponds to the harmonic mean of the estimated distances from the other two methods. As the physical geometry of a CME described by the fixed-phi and the harmonic mean method constitute as the extreme cases of their shape, a more generalised geometrical technique, called the SSEF, has been created by Davies et al. (2012). This method provides the potential to estimate the CMEs angular extent in the plane perpendicular to the viewing plane. For this reason the FP and HM form the limiting cases for the SSEF technique (Davies et al. 2012; Mostl & Davies 2012). Two more techniques have also been proposed in order to exploit the stereoscopic nature of STEREO; these are direct triangulations (Liu *et al.* 2010) and 'tangent-to-a-sphere' (Lugaz *et al.* 2010). However, as these two stereoscopic methods are not applicable to the proposed future missions of Solar Orbiter and Solar Probe, this paper concentrates with the single spacecraft techniques.

Time-elongation profiles for a given solar transient are often manually extracted from a time-elongation map (or 'J-map'). A J-map is created by extracting the intensities along a given position angle, PA (usually along the ecliptic) from a series of (often difference) images and plotting them as a function of elongation along the Y-axis and time along the X-axis (Davies *et al.* 2009). Theoretical error estimates have been made for such manual selections when the number of chosen points (N) is typically between 30 and 60 (Lugaz 2010). But as the size of N affects the associated uncertainty, a larger number of selected points are preferable. Savani et al. (2009) displayed a different technique than the manual selection by creating an automatic tracking procedure for the density feature found in the J-map. Although this method lowers the uncertainty by reducing the standard error of the residuals, as defined by Rouillard et al. (2010a) and Williams et al. (2009), in practice, the uncertainty is limited by the physical assumptions used to create the formulae. This uncertainty in the true direction of propagation and velocity is more clearly visible when the results from the point-P, FP and HM methods are compared (Lugaz 2010; Wood *et al.* 2010).

CMEs are often observed to have a characteristic three-part structure (Hundhausen 1993), interpreted and modelled by invoking a magnetic flux rope structure within the observed cavity (e.g. Cremades & Bothmer 2004; Thernisien *et al.* 2009; Wood *et al.* 2009). If the CME corresponds to a flux rope surrounded by a region of high electron density, then a single CME may exhibit two distinct features in the J-maps. This is because each track is associated to a density enhancement along a line of constant position angle (PA). The two features would represent the front and rear density enhancements. This implies that the elongation width of a CME can be extracted from a J-map (Lynch *et al.* 2010; Rouillard 2011). Lynch et al. (2010) showed for an event in June 2008 a double track in the J-map corresponded almost exactly to the edges of the magnetic flux rope signature and density

Page 5 of 30

enhancements as measured in situ. To convert this width in elongation angle to a physical width in distance requires a prior knowledge of the propagation direction, β. The heliocentric distance can be derived/calculated from the propagation direction and elongation value. By estimating and continuously tracking the radial width of a CME, for the first time, we are able to generate an empirical estimate for the expansion rate of a CME as it propagates between the Sun and 1AU. We display evidence that the CMEs expand linearly with heliocentric distance.

Many previous attempts to estimate the growth rate of CMEs have relied on a single CME size measurement made at a single position, allowing the expansion of radial widths from CMEs with heliocentric distance to be inferred only in a purely statistical sense, assuming no event-to-event variability. Such studies concluded that the radial widths of CMEs as a function of distance follow a power law (Bothmer & Schwenn 1994; Bothmer & Schwenn 1998; Gulisano et al. 2010). Savani et al. (2009) showed that the growth rate of a single CME can be estimated within the HI-1 field of view. The authors used two methods: the first was based on the radius of curvature of the CME's rear edge; in the second, the front and rear density features were manually followed for selected frames taken by the camera. Here, instead of tracking the density features manually within the images from the camera, we track the enhancements within J-maps. Therefore our method to estimate the radial width becomes easily applicable to the many events currently being observed by the STEREO HI instruments. This enables the results from the SolarStormWatch citizen science programme (SSW, www.solarstormwatch.com) to be used to extract statistical estimates for more properties of CMEs as they propagate out to 1AU.

In this paper, four individual CMEs identified by citizen scientists are investigated. It is hoped that these results provide a proof of concept, ready for when the large number of events produced by citizen scientists become statistically significant.

## 2. Solar Storm Watch project

This paper analyses four CMEs that were identified from the HI observations by citizen scientists working within the SolarStormWatch (SSW) project (Davis *et al.* 2012). SSW was first launched in March 2010 by a collaboration including the Royal Observatory Greenwich, the Rutherford Appleton Laboratory and the University of Oxford. It is part of the Zooniverse project[1] collection of citizen science projects (Fortson *et al.* 2011) which grew out of the Galaxy Zoo project (Lintott *et al.* 2008). One of the tasks within the SSW project involved the citizen scientists being instructed to find 'circular storms' and, as such, they were provided an example case study event investigated by Savani et al. (2009).

By searching for such 'circular storms', we were attempting to identify CMEs that were ejected end-on from the HI-1 cameras point of view. The geometry of an end-on CME would appear like the cross-section of a possible horizontal flux rope like structure (Cremades & Bothmer 2004). During the first nine month period post-launch of the SSW project, eleven circular-like CMEs were identified. In this paper we chose to investigate four of the clearest CMEs, which we define as: 1. being able to manually identify the contour of the CME edge; 2. possessing long tracks within J-maps (≳40º elongation). Figure 1 displays the four CMEs (February 15$^{th}$ 2008, August 07$^{th}$ 2008, May 09$^{th}$ 2009 and July 16$^{th}$ 2009 as defined by their entrance into the HI-1 field of view) at three different times during their propagation. The earliest frame for each CME displayed in figure 1 is from the COR-2 instrument, while the other two are from the HI-1 camera. Three of the events are observed in ST-A (COR-2 in the right-hand panel), and one in ST-B (COR-2 in the left-hand panel). The identification of the high electron density shell at the front and rear of the CME are indicated on the HI-1 images by the green and red arrows in the images, respectively. In these images the ecliptic plane roughly corresponds to the horizontal center line through the images.

As the HI and COR-2 cameras provide 2D information of the time evolution of CMEs, it is also possible to expand the study to observationally estimate the 2D aspect ratio of CMEs. Here we define the aspect ratio as the ratio of the vertical size to the radial width of a CME seen end-on, thereby quantifying the ellipticity of the CME's cylindrical cross-section (Savani *et al.* 2011a; 2011b). But while it is possible to individually track the vertical extent of a CME on a frame-by-frame basis, our aim is to produce a simple solution that is both easily repeatable and can be implemented into the SSW project. Savani et al. (2011a) showed that with simple geometry the vertical size can be estimated with two parameters; the total angular width subtended from the Sun ($\Omega$) and the heliocentric distance (R). The heliocentric distance of a CME is a variable that increases with time and can be extracted from the J-maps once the propagation direction of the CME (relative to the Sun-spacecraft line) has been estimated. The angular width can be considered to remain constant if we assume that CMEs propagate radially away from the Sun and we extrapolate that the measured angular widths remain constant from coronagraphs (St Cyr *et al.* 2000; Yashiro *et al.* 2004) to HI's field of view. This implies that a single measurement of Ω will still be required for each CME, but this can be incorporated into SSW or a similar crowd-sourcing project without a level of complexity that is too high for the citizen scientists. For CMEs that propagate in the plane of the sky, the total PA width of the CME is equal to its physical width with no projection effects. However, as many CME's do not propagate along the plane of sky both the radial width and the vertical size may be corrected for (Rouillard et al. 2009b).

---

[1] www.zooniverse.org

To date, the HI instrument has observed several hundred CMEs, and as we approach solar maximum the detection rate of CMEs by STEREO will increase. As such, citizen scientists are perhaps the best prepared to assimilate the data in a timely manner. Therefore it seems appropriate to engage citizen scientists to statistically estimate the average radial speed and propagation direction of CMEs. Active encouragement of the SSW programme has also begun to reveal significant strengths in providing forecasting capabilities of Earth impacting CMEs based on near-real time data. This is leading to preliminary results for space weather forecasting.

## 3. Analysis

In order to calculate the radial width of a CME, its direction of propagation and heliocentric distance must first be estimated. For the majority of CMEs the plane of sky assumption is no longer applicable for large elongations away from the Sun. Also, as the CME propagation direction deviates away from the plane of sky, the true angular width of the CME changes from the observed PA width seen on the plane of sky according to

$$\cos\frac{\pi}{2}-\lambda = \sin\delta\sin B0 + \cos\delta\cos PA\cos B0, \qquad (1)$$

$$\tan\phi = \sin PA\cos B0\sin\delta\cos B0 - \cos\delta\cos(PA)\sin B0. \qquad (2)$$

Here, λ and ϕ are the heliocentric latitude and longitude, respectively, with ϕ defined as positive to the right of the line of sight; δ is the angle of propagation away from the plane of sky from the observer (i.e. equal to (π/2) - β). $B_0$ is the latitude of the spacecraft. Further details of the geometry can be found in the appendix of Rouillard et al. (2009b) which is based on a point source of a small plasma parcel. As noted by the authors, the latitude must be considered with care depending on which quadrant the PA is evaluated in. In this paper, we estimate the latitudinal extent of the CME by subtracting the difference between the independently estimated northern and southern edges. It is worth noting that in this de-projection method the same δ is used for the top and bottom edge and not the same longitude, λ.

One method for determining the propagation direction of CMEs from observations extending to large elongations from the Sun is to analyse their time-elongation profiles extracted from J-maps. For each of the four CMEs, a J-map was created at a PA corresponding to the center of the propagating CME. Table 1 displays the PA along which the J-maps were taken as well as the angular width of the CME as observed on the plane of sky as a PA width. The difference between the PA widths and the projection corrected values vary according to the extent to which the CME propagation direction deviates away from the plane of sky. The correction in the angular width increases as the deviation away from the plane of sky increases. In majority of cases this is because the CME is travelling much closer to the camera than the plane of sky would suggest.

Table 1: Observational CME parameters.

| CME | Central PA (°) | PA Angular Width (°) | Projection-corrected Angular Width (°) | Average Beta β (°) | Heliocentric distance at HI-1 outer boundary (AU) |
|---|---|---|---|---|---|
| Feb 15 2008 | 90 | 35.0 | 33.8 | 102.6 | 0.51 |
| Aug 7 2008 | 103 | 30.0 | 25.7 | 61.1 | 0.41 |
| May 10 2009 | 265 | 35.0 | 30.3 | 60.9 | 0.41 |
| Jul 16 2009 | 93 | 25.0 | 15.6 | 39.4 | 0.45 |

Figure 2b displays a J-map created from a time sequence of running difference images. Each image is generated from the combination of COR-2, HI-1 and HI-2. A more detailed explanation for creating such J-maps can be found in Davies et al. (2009). Figure 2b displays the J-map along a PA of 265° for the May 9$^{th}$ event seen in ST-B with a false pink coloration to help identify the observed density enhancements. This map is then converted to a binary black-white figure by manually setting an arbitrary threshold of intensity (Savani *et al.* 2009), shown in figure 2c. The tracks corresponding to the front and rear edges are approximated by manually selecting a number of sample points along the J-map, N; where N is typically between 30 and 60 points (Davis *et al.* 2010; Lugaz 2010). The typical error associated with the manual selection procedure has been investigated by Williams et al. (2009) and Davis et al. (2010). It has been found that the error in the direction of propagation is usually of the order of 5°, but that uncertainty decreases as the maximum observed elongation angle increases. To minimise this uncertainty, observations to at least 30-40° elongations are desirable. More details on the fitting procedure can be found in Rouillard et al. (2010a) and Williams et al. (2009), however these authors investigate the FP method and further details on the HM method are discussed by Lugaz et al. (2010).

The uncertainty of the propagation direction can be reduced by increasing the number of sample data points, N. For this reason an automatic tracking technique was employed by Savani et al. (2009). This increased the number of sample points dramatically and therefore lowered the standard error parameter defined by Rouillard et al. (2010a). However even though this uncertainty was significantly lower, the realistic uncertainty is limited by the physical assumptions. The spread in results produced by the different methods of FP and HM expresses the true limitations of the physical assumptions (Lugaz 2010). Therefore by increasing the sample data points beyond ~60, the standard error of the residuals would no longer accurately reflect the true uncertainty in the propagation direction; for this reason, manual selection of between 30 and 60 points is employed in this study. Figure 3 displays the error estimate for both the front and rear edge of the May 2009 event as determined by the standard error of the residuals as expressed by Williams et al. (2009) and Rouillard et al. (2010a).

In this study we average the results between the FP and the HM technique for analysing the time-elongation profiles of CMEs. Although the geometry of the HM method is more physically realistic, the small CMEs investigated in this paper would likely represent a physical size in between the geometry presented by the FP and HM method. These techniques along with the SSEF technique are currently being used within the SSW project. The manually selected sample points are chosen on the binary image shown in figure 2c (red crosses). For each track (i.e. both the front and rear of the CME), the radial speed and propagation direction are retrieved using a non-linear optimisation routine (Nelder & Mead 1965) and displayed in table 2. However, error surface plots (see Figure 3) are also created to estimate the uncertainty in the optimal free parameters. For each CME, the propagation directions and radial speeds are averaged for the front and rear edge, and then between the FP and HM techniques (see table 2). Therefore the averaged propagation direction is assumed to be the direction of the CME nose, and the averaged velocity is assumed to represent the bulk flow at the center of the CME. However, we would expect the propagation direction to be the same for both front and rear edges, while for an expanding CME we would expect the speed at the front edge to be faster than the rear. Our results suggest that a discrepancy of ~10º in the propagation direction is not unfeasible and that the estimated speed of the front edge (average of ~300km/s) is preferentially slightly faster than the rear edge (~280km/s). These results are produced from only 4 slow streamer belt CMEs, therefore a larger sample size is required to verify these hypotheses and results.

Table 2: Results between the front and rear edge from the HM and FP technique.

| HM method CME | Beta, β | | | Velocity | | |
|---|---|---|---|---|---|---|
| | Front Edge (°) | Rear Edge (°) | Average (°) | Front Edge (km/s) | Rear Edge (km/s) | Average (km/s) |
| Feb 15 2008 | 108.2 | 120.9 | 114.5 | 301.9 | 304.6 | 282.2 |
| Aug 07 2008 | 64.5 | 61.3 | 62.9 | 326.6 | 296.2 | 311.4 |
| May 10 2009 | 57.7 | 72.7 | 65.2 | 295.3 | 269.0 | 282.2 |
| Jul 16 2009 | 46.2 | 39.4 | 42.8 | 284.9 | 269.9 | 277.4 |
| FP method CME | Front Edge (°) | Rear Edge (°) | Average (°) | Front Edge (km/s) | Rear Edge (km/s) | Average (km/s) |
| Feb 15 2008 | 86.8 | 94.6 | 90.7 | 284.9 | 279.6 | 282.2 |
| Aug 07 2008 | 60.0 | 58.5 | 59.3 | 325.6 | 297.5 | 311.6 |
| May 10 2009 | 52.1 | 61.0 | 56.6 | 293.2 | 257.0 | 275.1 |
| Jul 16 2009 | 33.7 | 38.3 | 36.0 | 297.4 | 280.1 | 288.7 |

From the manual selection, elongation values were linearly interpolated at constant increments of time along both the front and rear edges. From these interpolated values, the average elongation values were determined at any given time. These values were then assumed to be the center of the CME. Simple trigonometry was then used to generate a heliocentric distance for the center of the CME from the values of average elongation angle and average propagation direction. By a similar method, the radial width of the CME was estimated by deducing the heliocentric distance of the front and rear edge separately, and then subtracting the two results from one-another (see figure 4a). Again, the average propagation direction was used for the heliocentric estimates of both the front and rear edges. The radial widths of each CME as a function of heliocentric distance are displayed as coloured dashed curves in figure 4, along with best fit linear curves (in their respective solid colour). The best fit lines are estimated over the same range of heliocentric distances for all four CMEs, therefore the minimum and maximum heliocentric distances are limited by the faintest CME track.

The vertical size of a CME perpendicular to the direction of travel is estimated using COR-2 and HI-1 only. In HI-1, the four CMEs appear to propagate radially away from the Sun. We therefore assume radial propagation for distances beyond the field of view of HI-1, and convert the heliocentric distance of a CME and its total position angle (PA) width into a physical height/vertical size (Savani *et al.* 2011a). The total (projected) PA width for each of the four CMEs is estimated by manual inspection, and then converted to an angular width away from the plane of sky according to equation 1 and the geometry displayed in the appendix of Rouillard et al. (2009b), (see table 1). Note this is a single estimate for each CME after the angular extent appears to reach a maximum constant value (Yashiro *et al.* 2004). Therefore with this technique, the vertical size may be estimated out to the same heliocentric distance as the radial width, even though the density features begin to fade.

Using the linearly interpolated points, the radial width and the vertical size perpendicular to the radial line may be estimated over large heliocentric distances. Figure 4b displays the aspect ratio as a function of heliocentric distance for each CME (dashed colour lines), while the solid colour lines indicate the best fit linear curves. Once again, the best fit lines are limited to the distance range over which all four events could be observed (i.e. between 0.06-0.67AU). 0.93 AU was the largest heliocentric distance to be tracked for any of the four events studied (Feb 2008 CME). In this case the inner heliocentric boundary has been chosen by the Jul 2009 event, but in practice further statistical studies should not include smaller distances (≲0.1 AU) because the CME is unlikely to propagate at a constant speed, are often undergoing rapid expansion (Patsourakos et al. 2010) and can be considered to occur before the propagation phase in the CME's evolution. Figure 4 displays the rapid radial overexpansion compared to its rise above the Sun that occurs early on. The rapid decrease of the aspect ratio from ~3 to ~1 for the August 2008 event is consistent with the result found by Patsourakos et al. (2010). The approximately constant aspect ratio after ~0.5AU for three events (Feb08, May09 and Jul09) is consistent with the geometrical argument proposed by Savani et al. (2011a).

A significant purpose of this study is to show that this technique for studying the growth of CMEs could easily be applied to a large number of CMEs. However, since the CMEs under investigation are of relatively low speed, i.e. approximately 300km/s, and of the order of the nominal slow solar wind velocity, the derived results may be valid only to such 'swept up' type CMEs. The intention of comparing the growth rate of many CMEs, even with the large variability found between each event, is to highlight any similar underlying physical processes. Figure 5 displays the average of the linearly interpolated values of the radial width and the aspect ratio for the four events. The correlation coefficient for the average radial width and average aspect ratio curve are 0.959 and 0.996, respectively. These results reveal that the growth of CMEs, both in the propagation direction and perpendicular to that direction, is highly linear between the Sun and 1AU.

The coefficient value for the radial width linear best fit curve suggests that the expected expansion for slow streamer belt CMEs is constant at 7.1% of the bulk flow. A review of previous in situ studies by Forsyth et al. (2006) has showed that the expansion rate may change with heliocentric distance but is not strongly dependant on the distance. Their study shows that the expansion rate can be considered to be constant at approximately 10% of the bulk flow. We suggest the discrepancy may originate from the difference in expansion behaviour between slow and fast CMEs. If slow CMEs expand at a constant rate and fast CMEs follow a power law, then a statistical survey that does not distinguish between the two groups may result in an expansion rate that is weakly dependent on distance.

The estimate of the average aspect ratio at 1AU is 3.4; this is consistent with geometrical estimates (Savani et al. 2011a) and in situ measurements (Russell & Mulligan 2002; Savani et

Page 13 of 30

al. 2011b). The constant to the linear best fit curve (0.9) indicates that at the initiation phase of a CME the radial width is greater than the latitudinal extent. This is consistent with overexpanding observations seen in coronagraphs.

In this paper, a detailed analysis for remote and in situ comparisons was considered beyond the scope of this work. The events in this paper were chosen on the basis of those identified by the SSW project and not by investigating an ideal case study for remote to in situ tracking (e.g. Mostl et al. 2009; Mostl et al. 2010; Davis et al. 2011; Rouillard et al. 2009a; Rouillard et al. 2009b; Rouillard et al. 2010a; Rouillard et al. 2010b). For the four events investigated, they either did not travel towards a spacecraft (Feb 2008) or the in situ signatures did not clearly define a large magnetic CME structure (Aug 2008, May 2009 and Jul 2009). For August 2008, the event is travelling slightly below the ecliptic, such that any in situ detection would likely display measurements of the CME's outer northern edge. Therefore this event is far from an ideal case. The July 2009 event displays a very small event with a duration of about six hours (between 03:00 UT and 09:00 UT on 22 July 2009). Within this short time interval the rotation of the magnetic field and a weak bi-directional electron signature may represent the passing of a transient within the in situ data at L1, however the event is small and not an idealised magnetic flux rope. The May 2009 event is estimated to propagate close to the L1 point (~14º west of Earth). The ACE and WIND data displays density enhancements at the expected positions ahead (03.50UT 14$^{th}$ May) and behind (14.45UT 14$^{th}$ May) the CME (see Figure 2b). However as the nose of the CME is not predicted to arrive at L1, and that the clearest double J-track occurs 8º below the ecliptic, the magnetic signature does not show a clear start and end of the event. The velocity profile within the density interval does not display the ideal linearly decreasing speed profile often observed in large, fast moving expanding magnetic flux ropes. Instead, the velocity profile displayed in Figure 2b shows an approximate steady speed of ~340km/s, which is both in sharp contrast to the background wind speed ahead and behind the CME. This corroborates with figure 4a which displays that the event appears to maintain a more constant radial width after 0.7AU.

From the four events studied in this paper, two appear to be significantly expanding (Feb 2008 and May 2009) in the radial direction while the other two expand at a much slower rate between 0.1 AU to 0.7 AU. These results express the inherent variability of observed CMEs as seen remotely and in situ. As such, the average expansion rate may in fact be incorrect for any one particular event. Therefore the average expansion rate would be better studied in a statistical manner with a larger number of events so that the central limit theorem may be implemented. Future studies from the Solar Storm watch project may help in estimating the probability of encountering events with either small or very large expansion speeds. In addition, the flux-rope like nature of the May 2009 event as seen in COR2 and HI-1 is not expressed in the in situ magnetic field data or as a linearly decreasing speed profile. This event suggests that there may be a higher probability of CMEs possessing a flux rope structure than are currently observed in situ.

To help validate our claims for the expansion rate, an investigation by Lynch et al. (2010) showed that the well studied event from the 1$^{st}$ June 2008 also displays a double J-track. The authors estimated the current expansion as measured in situ at ST-B as ~6% of the bulk flow, which is in reasonable agreement with our analysis. However, our results for this event would have generated an expansion of 8% from the data of the double J-track at 0.7AU. This suggests that the mathematical formulism of a constant expansion rate may overestimate the true expansion rate for larger and faster events. Under these circumstances the interaction between a CME and the solar wind may decrease the expansion as it propagates into the heliosphere. Therefore this may indicate faster CMEs could be more appropriately measured with a power law.

## 4. Discussion and Conclusions

In this paper we have investigated four CMEs observed by STEREO spacecraft that were identified by citizen scientists from the SolarStormWatch (SSW) project. These CMEs propagated at relatively low speeds (i.e. below 300km/s) and can be considered as being embedded in the nominal slow wind. The average results between the fixed phi and the harmonic mean fittings of the time-elongation profiles from J-maps haves been implemented as the most appropriate method to deduce the radial speed and propagation direction. This has the advantage of being easily applied to many events seen in wide-angle images, such as from the HI cameras. By tracking the density features observed at the front and to the rear of a CME, the radial width of the CME can be routinely estimated as a function of heliocentric distance, up to R~0.7AU in this case. By first measuring the CME's projected PA angular width once it converges to a constant value in the coronagraphs (St Cyr *et al.* 2000; Yashiro *et al.* 2004), and then assuming this size is maintained throughout the HI field of view, the vertical size of a CME perpendicular to the radial width may be estimated as a function of R. This leads to a repeatable procedure to estimate the 2D structure of many CMEs between the Sun and the Earth. The tracking of the radial width with heliocentric distance (R) enables estimates of the average expansion rate (A) between the solar surface and R (i.e. over its entire lifetime of propagation) to be generated. We show that the mean expansion rate for the four CMEs is 7.1% of the bulk flow. The propagation effects are analysed after the CME reaches ~0.1AU, before which the CME undergoes a more rapid expansion which indicates that CMEs are still within a stage of development below ~0.1AU (Robbrecht et al. 2009; Patsourakos et al. 2010).

An idealised estimate for the cross section of an average CME from the four events studied is schematically shown as the blue shaded region in Figure 6. This schematic is not intended to represent an accurate and detailed structure of a CME, but to display a large scale estimate by concentrating on four definitive 'vertices' of the CME – the top, bottom, front and rear edge. The radial width and aspect ratio from Figure 5 are used to generate the 2D shape. The red contour displays the 2D structure by assuming the radial width follows the in situ estimates from the power law derived by Bothmer et al. (1998), while the vertical size is estimated from the average PA angular width from a LASCO survey (St Cyr *et al.* 2000). The average radial width of the four CMEs studied in this paper is thinner than is suggested by the average in situ measurements. Often magnetic flux estimates of a CME are deduced by assuming a circular cross-section from the in situ measurements (Owens *et al.* 2008). For this reason, the pink circular shaded region represents the size and shape of an assumed constant alpha flux rope as defined by in situ measurements (Bothmer & Schwenn 1994; 1998).

In this work, we assume that the regions of the CME can be demarcated by the location of more intense line-of-sight integrated Thomson scattered white light. These regions correspond to areas of higher total electron content along the line of sight. By comparing such imaging observations of CMEs with in situ measurements, these high density regions are found to surround the CME magnetic structure (Davis *et al.* 2009; Mostl *et al.* 2009; Rouillard *et al.* 2009a; Rouillard *et al.* 2010b). As such, by defining the CME size from these features in our analysis, we may estimate a slightly larger CME structure than the true size. However, any possible discrepancy is likely to be a systematic error over all heliocentric distances and all events. Therefore, the effect will be on the absolute size of the CME and not the relative growth rate; meaning the gradient of the curves in figure 5 should remain unaffected. In situ measurements for the May 2009 event displays a speed of about 340km/s for a duration of 11 hours. This correlates to approximately 0.09AU in radial width. However, the remote sensing tracks would suggest a value of 0.13AU, which is consistent with the idea that the CME is travelling below the ecliptic and therefore the spacecraft at L1 would cut through a smaller cross-section. Although, this small discrepancy might also be, in part, due to our observations overestimating the radial width of CMEs.

In this paper, details of the in situ measurements for the four events have not been thoroughly investigated as the events either did not travel towards a spacecraft (Feb 2008, Jul 2009) or the in situ signatures did not clearly define a CME structure (Aug 2008, May 2009). Future work should consider investigating the expansion speed by directly comparing STEREO remote sensing to a clear in situ magnetic flux rope with MHD simulations, or possibly by investigating the expansion rate of a CME measured in situ by multiple spacecraft at different heliocentric distances. Perhaps an emphasis on distinguishing the properties between fast and slow CMEs may shed further light onto whether the expansion speed remains constant or follows a power law.

For a case study event, Lynch et al. (2010) compared in situ measurements to the density features of a CME observed by HI. The authors showed that any discrepancy in the true radial width is minimal. Their investigation assumed a circular shape propagating in the plane of sky for COR-2 and studied individual pixels on a frame-by-frame basis in HI running difference images. Also, geometrical arguments used for estimating the vertical extent of CMEs and therefore the aspect ratio has been dependant on cross-sections of a 2-D shape away from the plane of sky. This may not be entirely accurate for a 3-D structure. Further understanding of the projection effects from a 3-D CME is required to understand any disagreement. However, Savani et al. (2010) demonstrated that reasonable agreement between observations of a distorted CME away from the plane of sky, and estimated structures of a modelled CME in the plane of sky may still be produced. Using in situ data, Lynch et al. (2010) estimated the average bulk flow of the CME to be 390 km/s with an expansion speed of 24.5km/s. This implies an expansion rate of 6.3% which is very consistent to the average of the four events we studied (7.1%).

The HM technique was used to estimate the radial speed and propagation direction of four selected CMEs from time-elongation profiles extracted from J-maps. These J-maps were generated along lines of constant PA which followed the center of the CME. Estimates of β and Vr, using an optimisation routine, were made for both the front and rear edges of each CME. It is expected that the radial velocity would be different for the two edges if the CME is observed to expand. But the propagation direction, often considered to be the position of the CME nose should be the roughly same for the two features. Currently, as the accuracy of both the leading and rear edge estimates may be considered to be similar, no preference can be given to either. For this reason an average of the propagation directions was implemented in this study. As the accuracy of the propagation direction improves for larger elongation ranges, further statistical studies of many CMEs may consider the benefits of implementing 'a preferential system' of selecting one of the propagation directions. Future surveys over a large number of events may find a statistically significant difference between the radial speeds of the front and rear edges.

The direct measurements of the radial width for two events (shown by the dashed lines in Figure 4 for the May09 and Feb08 events) show a small break in the curves at ~0.45AU. Thus it would appear that the curve for each CME might be better represented by two separate linear best fits, one for heliocentric distance of 0.1-0.45AU and the other for 0.45-0.65AU. The second half of the propagation shows a dramatic increase in the growth rate of the radial size. However, closer inspection into the elongation position reveals that this represents the boundary between the HI-1 and HI-2 cameras. This may simply be due to a different cadence on the cameras, or a combination of a different response between the cameras due to their different pixel sizes, and a greater image blurring from the CME moving into HI-2 which has a longer exposure time. However as each elongation bin within a Jmap incorporates many pixels, the effects of pixel size and image blurring is likely to be negligible. It is worth noting that an alternative interpretation has been brought forward for the changing expansion rate centered on ~0.5AU. Bothmer and Schwenn (1998) suggest it may be due to the approximate position where the density within a CME drops to below the average background solar wind. Therefore the interaction and dynamics between solar wind and CME would change.

Figure 5 reveals that the radial width and the 2D aspect ratio averaged over four CMEs have a linear dependency with heliocentric distance; this is different to the estimated results from in situ measurements. The radial width of CMEs as a function of heliocentric distance has previously been fitted to power laws (e.g. Bothmer & Schwenn 1994; 1998; Liu *et al.* 2005; Wang *et al.* 2005). However these statistical studies rely on a single point measurement per CME, and therefore observe a large variability in the results due to the variety of CME sizes. The technique demonstrated here tracks individual CMEs in a similar method to that of Savani et al. (2009); these authors estimated the size from the radius of curvature of the rear edge on a frame-by-frame basis. But by tracking a larger number of events, future studies may be able to justify a power law behaviour for CMEs.

By inspection of figure 4, the data shows that even though the absolute values of the radial width and aspect ratio may differ, linear best fits for each event is appropriate. This demonstrates that a similar underlying physical process may be affecting the size of slowly propagating CMEs. Whereas faster CMEs are more likely to interact with the solar wind by being compressed from the front; this would decrease the expansion speed as a function of distance and therefore more likely to produce a power law behaviour. With the onset of more wide-angle cameras proposed on future missions, future studies should begin to move away from statistical one-dimensional studies (e.g. Bothmer & Schwenn 1998) and start to incorporate the 2D structures that are now being observed further into the heliosphere.

Future work should consider taking the large catalogue of CME events from wide angle cameras and statistically analyse the 2D structure using the techniques above. This is so that the average growth rate of CMEs may be more reliably estimated and to investigate any differences between slow and fast CMEs. The SolarStormWatch project may prove to be an invaluable tool in the cataloguing process for not only the STEREO mission but also the possible future missions of Solar Orbiter and Solar Probe Plus.

**Acknowledgments**. NPS would also like to give my deepest appreciation to Japan society for the promotion of science (JSPS) and JSPS London for the award of the (short-term) postdoctoral fellowship and the fantastic opportunity to work within Japan. This research was also supported by the NASA Living With a Star Jack Eddy Postdoctoral Fellowship Program, administered by the UCAR Visiting Scientist Programs and hosted by the Naval Research Laboratory. NPS would like to thank N. Sheeley for useful discussion, and the science team and citizen scientists from the SolarStormWatch project and the Royal Observatory Greenwich.

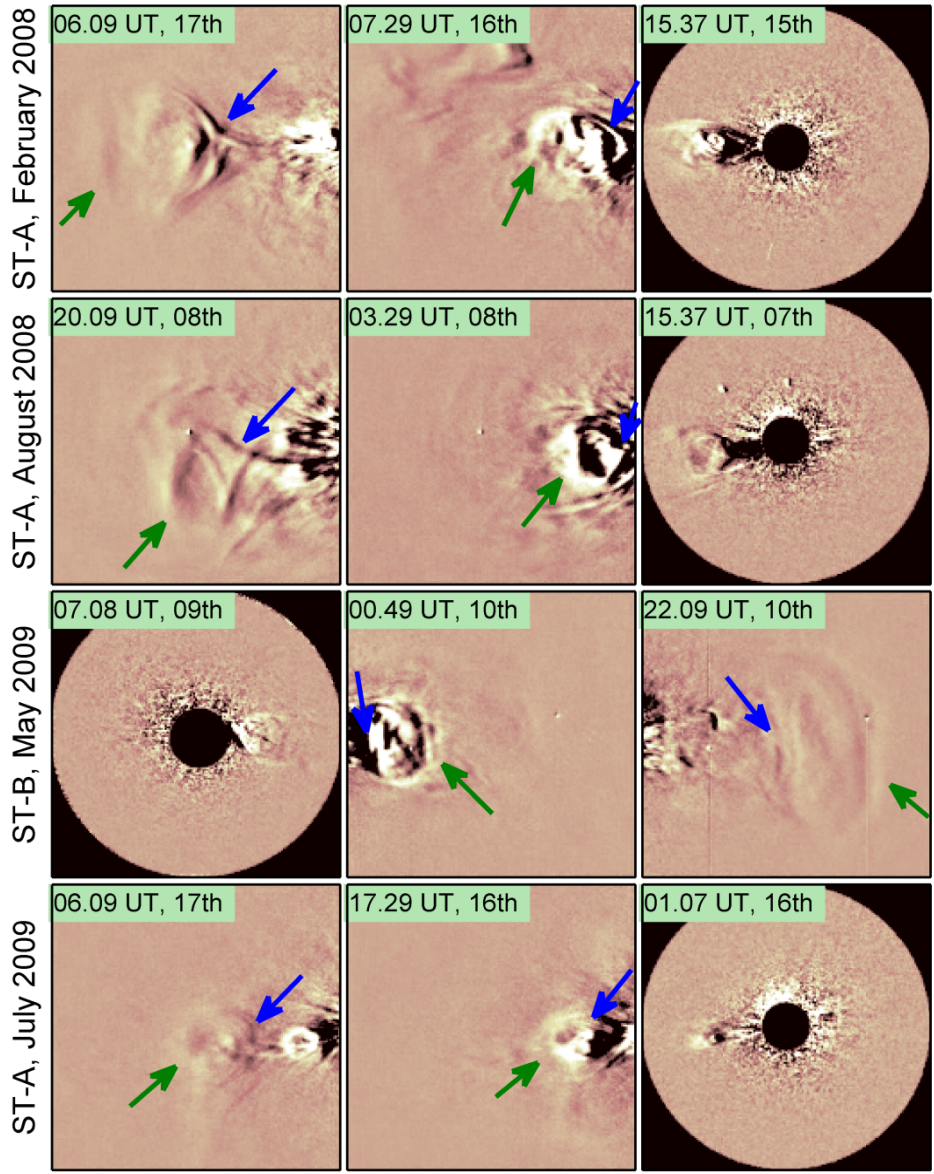

Figure 1. Four CMEs used in this study are individually displayed on each row. Three frames in time display the CME structure evolving from COR-2 to the outer edge of HI-, from right to left for STEREO-A, and left to right for STEREO-B. The field of view of COR-2 extends to 4° elongation away from the Sun, while the HI-1 field of view is 4-24°.

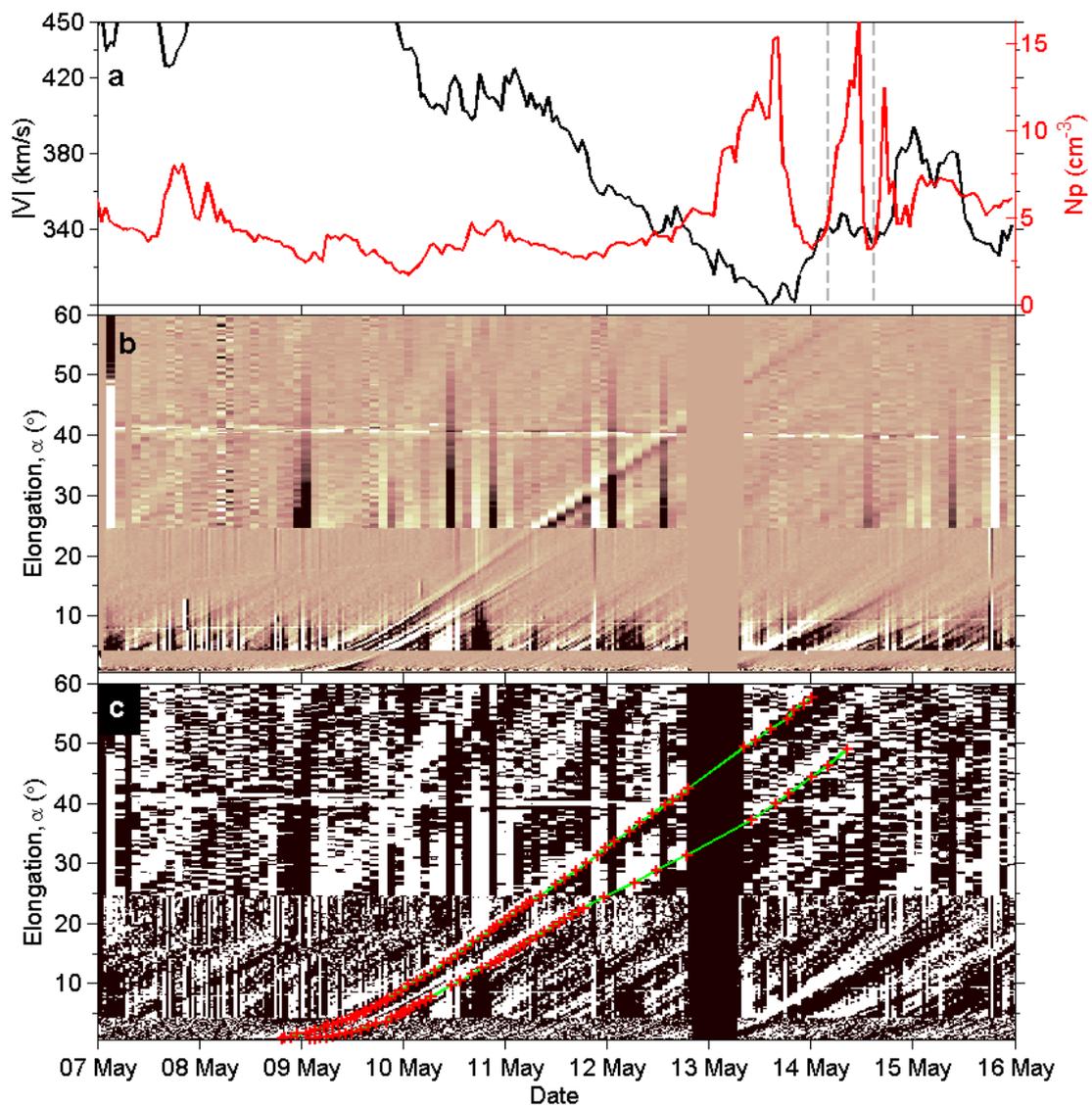

Figure 2. a) In situ measurements of solar wind speed (black) and density (red) from the WIND spacecraft. The dashed vertical lines indicate the start and end positions of the CME defined by the J-map tracks. b) J-map for May 09$^{th}$ 2009 event. c) Displaying manually selected data points for both the front and rear edge using a black and white image, which has been created by setting an arbitrary threshold value. The green line is created by linearly interpolating the selected points at regular time intervals. The interpolated values of the green line are then used in the subsequent analysis.

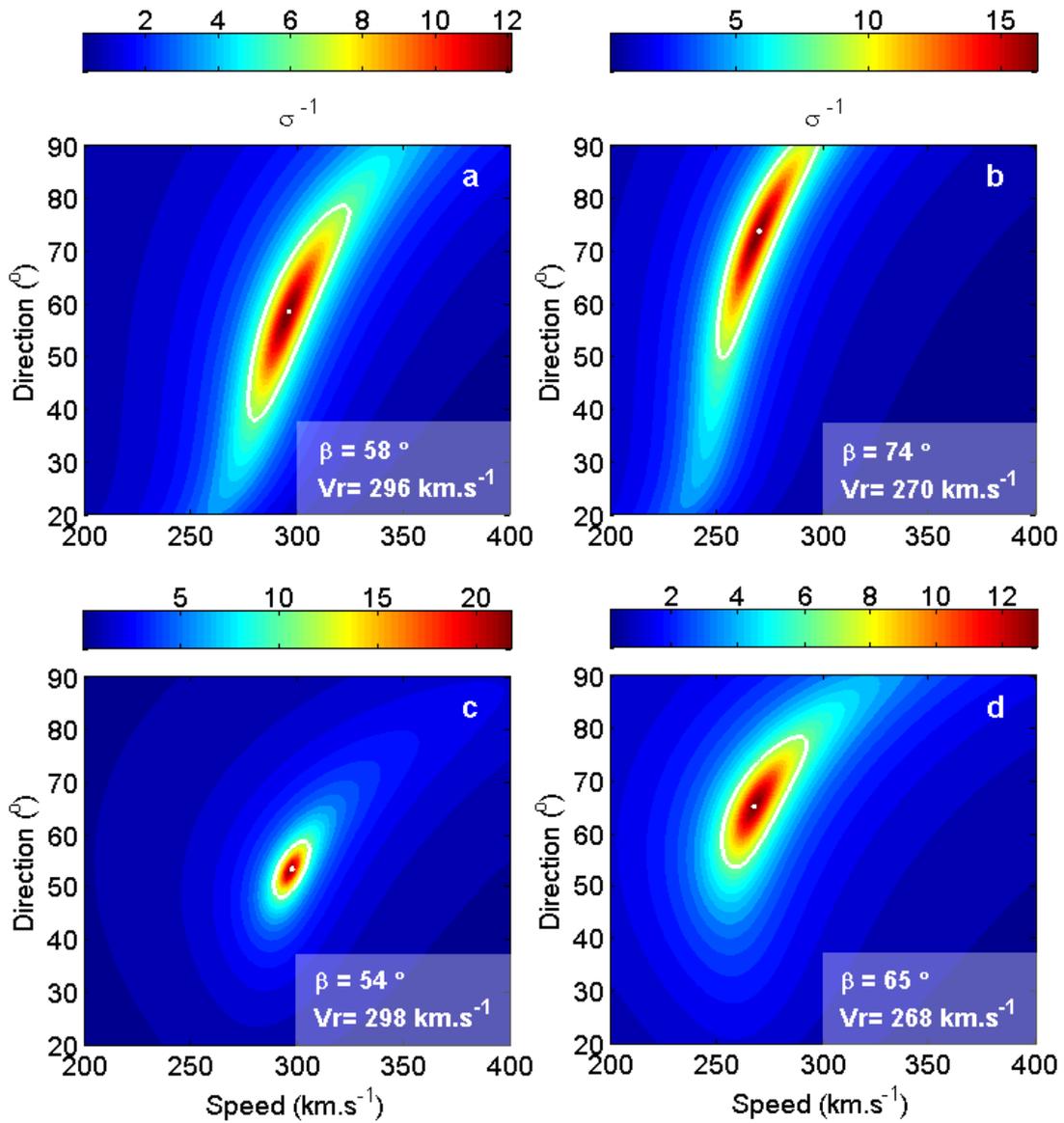

Figure 3. The inverse of the standard error of the residuals ($\sigma^{-1}$) as defined by Williams et al. (2009) and Rouillard et al. (2010a). $\sigma^{-1}$ are displayed as a function of the two free variables; radial speed (Vr) and propagation angle (β). a) and c) Shows the optimised parameters for the front edge in the J-map of Figure 2; b) and d) shows the rear edge. The top and bottom rows are for the HM and FP technique respectively.

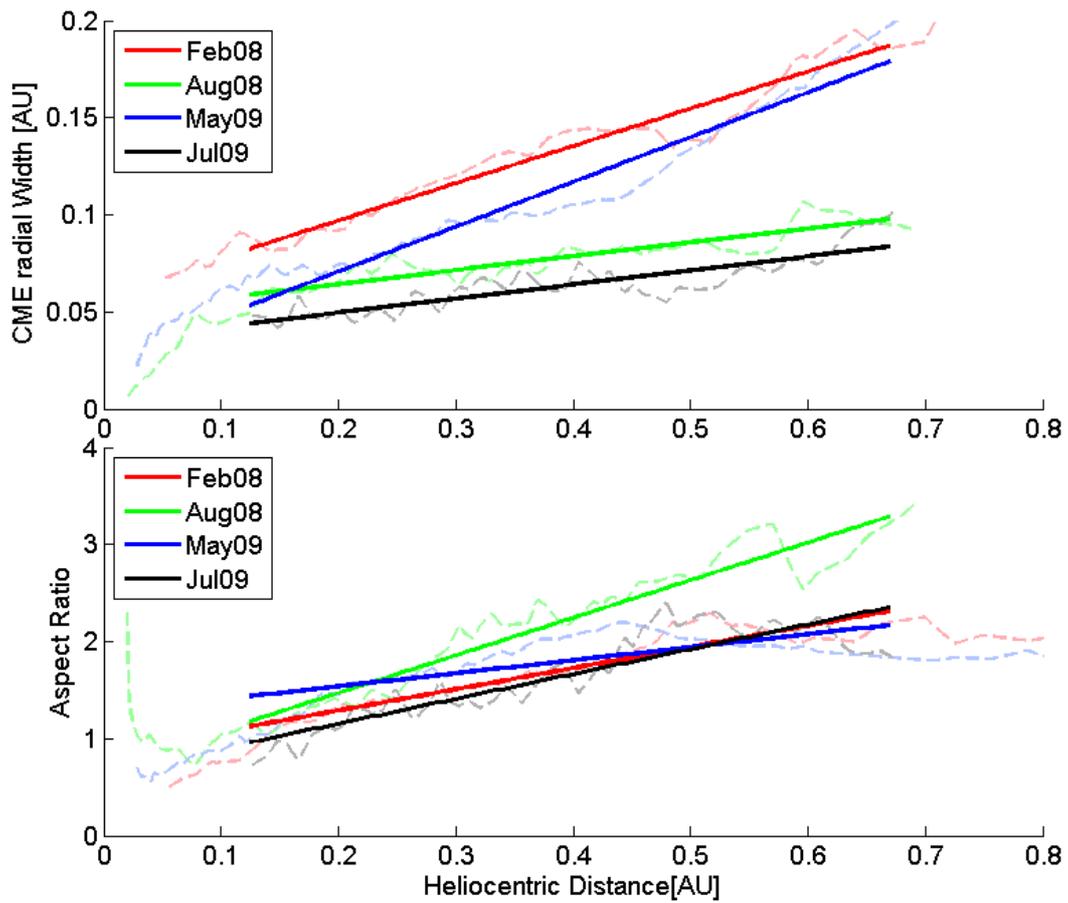

Figure 4. Top) displays the radial width of the four individual CMEs as they propagate anti-sunward. Bottom) Displays the aspect ratio of each CME as a function of heliocentric distance. The vertical extent of the CME is estimated from observed total PA width and estimated heliocentric distance. The dashed curves are taken from the interpolated elongation widths and converted into distances in AU, depending on the direction of propagation with respect to the Sun-spacecraft line. The solid lines are the best fit linear curves between 0.06AU and 0.67AU.

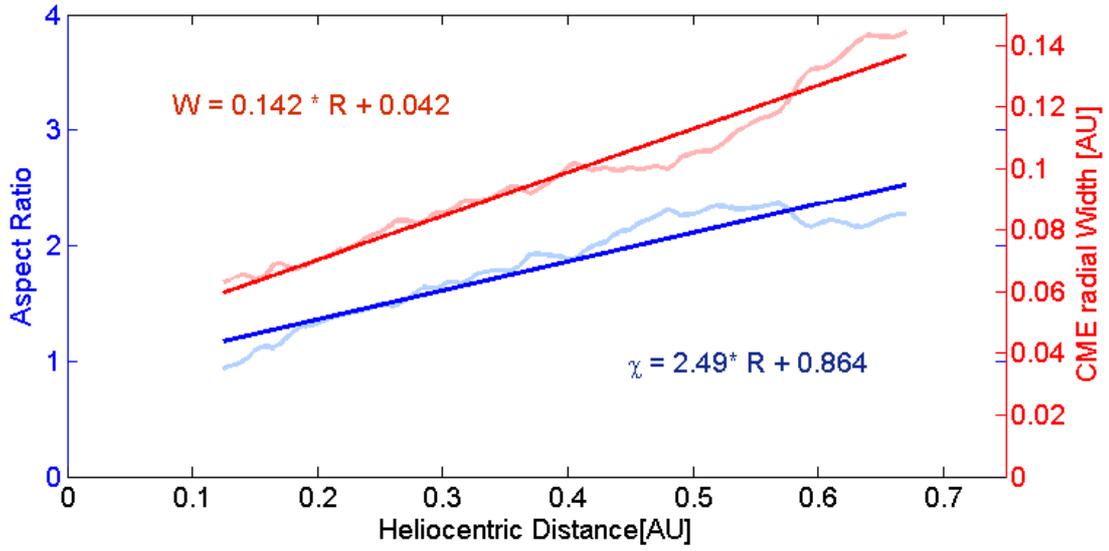

Figure 5. Both the growth in the radial width (red) and aspect ratio (blue) are displayed as a mean of all four events between 0.06AU and 0.67AU. The linear best fit curves are included with their respective equation.

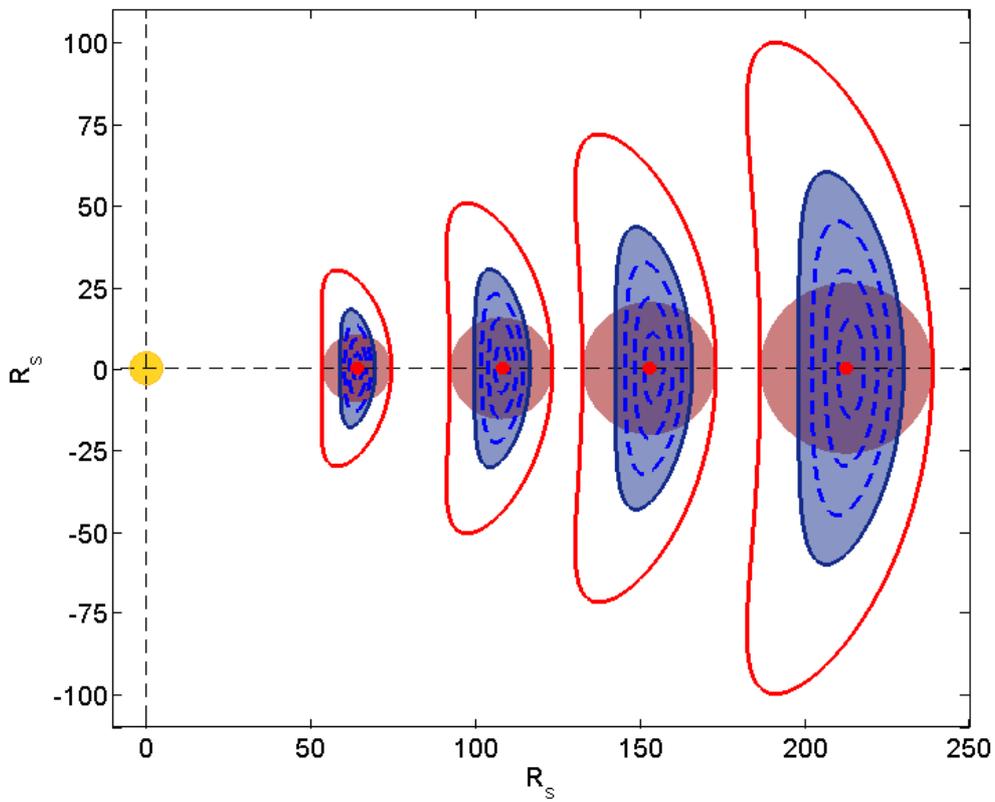

Figure 6. Displays a schematic that compares the different shapes assumed by various measurements and models. The red contour displays the 2-D structure by assuming the radial width follows the in situ estimates from Bothmer et al. (1998)

power law, while the vertical size is estimated for the average projected angular width from a LASCO survey (St. Cyr, 2000). The pink circular shaded region is the size and shape of an assumed constant alpha flux rope as defined by in situ measurements (Bothmer et al. 1994, 1998). The blue contour and associated shaded region is deduced from the HI instrument discussed in this work. The angular extent is the averaged projected plane of sky value, as these are the cross sections that would be observed for ideal CMEs propagating along the plane of sky.